\documentclass[10pt,conference,letterpaper]{IEEEtran}
\usepackage{cite}
\usepackage{graphicx}
\usepackage{psfrag}
\usepackage{subfigure}
\usepackage{diagbox}
\usepackage{color}

\usepackage{url}
\usepackage{amsmath}
\usepackage{array}
\usepackage{graphics}
\usepackage{epsfig}
\usepackage{mathtools}
\usepackage{amsbsy}
\usepackage{amssymb}
\usepackage{amsthm}
\newtheorem{theorem}{Theorem}

\newtheorem{lemma}{Lemma}
\newtheorem{remark}{Remark}

\usepackage{framed}
\usepackage{algorithm}
 
\DeclarePairedDelimiter\ceil{\lceil}{\rceil}
\DeclarePairedDelimiter\floor{\lfloor}{\rfloor}

\IEEEoverridecommandlockouts

\begin{document}

\title{Reliability and Local Delay in Wireless Networks: Does Bandwidth Partitioning Help?\vspace*{-2mm}}
%
%
%
\author{\IEEEauthorblockN{Sanket S. Kalamkar}
\IEEEauthorblockA{INRIA and LINCS\\
Paris, France\\
E-mail: sanket.kalamkar@inria.fr}
\vspace*{-6mm}}

\maketitle

\begin{abstract}
This paper studies the effect of bandwidth partitioning (BWP) on the reliability and delay performance in infrastructureless wireless networks. The reliability performance is characterized by the density of concurrent transmissions that satisfy a certain reliability (outage) constraint and the delay performance by so-called {\em local delay}, defined as the average number of time slots required to successfully transmit a packet. We concentrate on the ultrareliable regime where the target outage probability is close to $0$. BWP has two conflicting effects: while the interference is reduced as the concurrent transmissions are divided over multiple frequency bands, the signal-to-interference ratio (SIR) requirement is increased due to smaller allocated bandwidth if the data rate is to be kept constant. Instead, if the SIR requirement is to be kept the same, BWP reduces the data rate and in turn increases the local delay. For these two approaches with adaptive and fixed SIR requirements, we derive closed-form expressions of the local delay and the maximum density of reliable transmissions in the ultrareliable regime. Our analysis shows that, in the ultrareliable regime, BWP leads to the reliability-delay tradeoff. 
\end{abstract}
\IEEEpeerreviewmaketitle

\section{Introduction}
\subsection{Motivation}
The performance in a wireless network depends critically on the allocated bandwidth. For instance, in an interference-limited wireless network, the data rate $R$ over a link depends on bandwidth $W$ as $R = W\log_2(1+\mathsf{SIR})$, where $\mathsf{SIR}$ is the signal-to-interference ratio for the link under consideration. Often, an operator divides available bandwidth into smaller frequency bands, and the users select one or more sub-bands for their communication. Such a bandwidth partitioning (BWP) affects the data rate. In particular, suppose the entire bandwidth $W$ is split into $N$ sub-bands and each transmitter selects one sub-band for transmission. In this case, the supported data rate is $\frac{W}{N}\log_2(1+\mathsf{SIR})$.

Increasing the number of sub-bands $N$ has two conflicting effects. On the one hand, it reduces the interference since the concurrent transmissions are divided into $N$ bands. On the other hand, to maintain the target data rate at $R$, the SIR requirement, which is $2^{NR/W}-1$, grows with $N$. This exponential rise in the SIR requirement could potentially hurt the SIR-based performance in wireless networks. For example, commonly in wireless communication, a packet transmission is considered successful if the receiver achieves the minimum SIR. Hence, any increase in the SIR requirement reduces the probability of a successful packet transmission, which often quantifies the reliability in wireless networks. But, if the SIR requirement is to be kept the same, splitting the bandwidth into $N$ sub-bands results in $N$-fold reduction in data rate. This means that the time required to send a fixed size packet increases $N$-fold, leading to a higher transmission delay. Taking these two conflicting aspects into account, this paper attempts to answer the following question: Does bandwidth partitioning help improve the reliability and the delay performance in infrastructureless wireless networks?

\subsection{Reliability and Delay}
We focus on {\em outage} scenarios where a successful transmission needs the received SIR to be larger than a threshold. That is, the reliability is determined by the events of SIR exceeding some threshold. In this context, subject to an outage (reliability) constraint, the maximum number of successful transmissions per unit area of the network may be thought as an indicator of the number of users that can reliably be accommodated in the network. This density of reliable transmissions is an extremely useful performance metric in modern wireless networks because it sheds light on the questions of network densification under strict reliability constraints. 

The delay is characterized by the amount of time required to successfully transmit a message. Specifically, we quantify the delay as the average number of time slots needed for a transmitter to successfully transmit a packet to its next-hop receiver, which is called the \textit{local delay}~\cite{Baccelli_2010, Haenggi_Local}.\footnote{In this paper, we focus on fully backlogged transmitters that always have packets to transmit. Hence, the local delay is the transmission delay and not the queuing delay.}\vspace*{1mm}

\subsection{Contributions}
\begin{itemize}
\item[1)] For infrastructureless wireless networks, depending on whether the SIR requirement is changed or not with the number of sub-bands $N$, we analyze and compare two approaches using stochastic geometry to study the effect of BWP on the density of reliable transmissions and the local delay.
\item[2)] We obtain simple closed-form expressions of the local delay and the maximum density of reliable transmissions in the ultrareliable regime.
\item[3)] We show that, in the ultrareliable regime, optimizing the number of sub-bands with the sole aim of maximizing the density of reliable transmissions leads to infinite local delay for both the adaptive and fixed SIR requirement approaches. 
\end{itemize} 

\subsection{Related Work}
From the stochastic geometry approach, the works in~\cite{Jindal_2008} and \cite{Zhong2014} are perhaps the most relevant. The goal of~\cite{Jindal_2008} coincides with one of our goals---the use of BWP to maximize the density of transmissions that meet an outage constraint in {\em ad hoc} networks. But, \cite{Jindal_2008} applies the outage constraint at the {\em typical} link which represents the average performance of all links in a snapshot of the network. Thus the outage constraint is not applied at each individual link. In contrast, our framework applies the outage constraint at \textit{each} individual link and hence actually calculates the density of transmissions that meet an outage constraint. Also, \cite{Jindal_2008} does not study the effect of BWP on delay, while our work does. Finally, \cite{Jindal_2008} considers only the approach where the SIR requirement is changed with $N$ to keep the data rate the same. The work in~\cite{Zhong2014} focuses on the local delay analysis with BWP for the fixed SIR requirement approach and does not consider the reliability aspect. 
 
In~\cite{Mehanna2013}, for two heterogeneous networks coexisting together, a comparison between spectrum sharing and spectrum splitting is studied in terms of the average spectral efficiency. A recent work~\cite{Lu_2019} focuses on BWP for frequency-selective channels with an aim to maximize the total density of concurrent transmissions subject to the outage constraint at the typical link. In~\cite{soc_2017}, the density of reliable transmissions is maximized for ALOHA-based uncoordinated wireless networks without BWP. There are also several works on BWP for coordinated random wireless networks such as cellular networks (see~{\em e.g.},~\cite{Stefanatos_2014, Saha_2018}) and wireless networks with deterministic node locations (see~{\em e.g.},~\cite{Yeung}). 


\section{System Model}\vspace*{-1mm}
\subsection{Network Model}
We consider a network model where the transmitter locations follow a homogeneous Poisson point process (PPP) $\Phi \subset \mathbb{R}^2$ of intensity $\lambda$. Each transmitter has an associated receiver at unit distance in a random direction. This model is known as the Poisson bipolar network and is widely used to study infrastructureless networks such as device-to-device (D2D) and machine-to-machine (M2M) networks~\cite{bb_book}. Due to the stationarity of the PPP, it is sufficient to focus on the reference link between the receiver located at the origin and its associated transmitter. In other words, we condition on the transmitter located at $x_0 \in \Phi$ whose receiver is located at unit distance at the origin, {\em i.e.}, $\|x_0\| = 1$. The link between the transmitter at $x_0$  and its receiver is called the typical link after averaging over the PPP in the sense that this link has the same statistical properties as all other links in the network. 


The network operates in a time-slotted manner where the time is divided into slots of equal duration. Each packet is of a fixed size, and it takes exactly one slot for a packet transmission if the entire bandwidth $W$ is used. We focus on the interference-limited regime.\footnote{The results in this paper can easily be extended to scenarios with both interference and noise taken into account.} A transmission is considered successful if the received SIR exceeds the predefined threshold $\theta$. If successful, a transmitter can send information at spectral efficiency $\log_2(1 + \theta)$ bits/s/Hz. Let $R$ denote the data rate of a transmitter when it uses the entire bandwidth $W$ for its transmission, {\em i.e.}, $R = W\log_2(1 + \theta)$, which corresponds to the SIR threshold $\theta = 2^{R/W}-1$. 

All transmitters transmit at unit power and always have packets to transmit. A transmission over distance $r$ is subject to the path loss as $\ell(r) = r^{-\alpha}$, where $\alpha > 2$ is the path-loss exponent. We assume Rayleigh fading, where the channel power gain is exponentially distributed with mean $1$. Let $h_{k, x}$ denote the channel power gain between the transmitter $x \in \Phi$ and the typical receiver at the origin in $k$th time slot. 
\subsection{Performance Metrics} 
\label{sec:perfor_metric}
We focus on the performance metrics that are based on the SIR threshold model discussed in the previous subsection.
\subsubsection{Density of Reliable Transmissions}

For a stationary and ergodic point process $\Phi$, the density of reliable transmissions, {\em i.e.}, the density of concurrent transmissions that meet the outage constraint, is defined as
\begin{align}
\lambda_{\rm s} \triangleq  \lambda \bar{F}(\theta, \varepsilon), \quad \theta > 0,~\varepsilon \in [0, 1],
\label{eq:dens_suc}
\end{align}
where $\bar{F}(\theta, \varepsilon)$ is the fraction of links that achieve an SIR of $\theta$ in each realization of the point process $\Phi$ with probability at least $1-\varepsilon$. Alternatively,  $\bar{F}(\theta, \varepsilon)$ represents the fraction of reliable links in each realization of $\Phi$. Note that $\varepsilon$ corresponds to the target reliability or the target outage probability. The intensity $\lambda$ of $\Phi$ affects $\lambda_{\rm s}$ positively as well as negatively: an increase in $\lambda$ increases the density of links but reduces the fraction of links that meet the outage constraint due to the increased interference. Hence there exists an optimal $\lambda$ that maximizes the density of reliable transmissions, which we shall calculate later in Theorem~\ref{thm:max_dens}. 

For an ergodic point process-based model, the fraction $\bar{F}(\theta, \varepsilon)$ is the SIR meta distribution (MD)~\cite{martin_meta_2016}, which is a fine-grained performance metric that allows one to calculate per-link reliability in each realization of the network. The SIR MD is defined as the distribution of the conditional success probability $P_{\rm s}$ given a realization of $\Phi$. Mathematically,
\begin{align*}
\bar{F}(\theta, \varepsilon) \triangleq \mathbb{P}^{!}_{o}(P_{\rm s}(\theta) > 1-\varepsilon), 
\end{align*}
where $\mathbb{P}^{!}_{o}(\cdot)$ denotes the reduced Palm probability conditioned on the typical receiver at the origin $o$ and its associated transmitter being active. The link success probability $P_{\rm s}$ (alternatively, the link reliability) is given by
\begin{align*}
P_{\rm s}(\theta, \Phi) \triangleq \mathbb{P}(\mathsf{SIR} > \theta \mid \Phi),
\end{align*}
where the averaging is done over the fading and the channel access scheme.  

Letting $\varepsilon \to 0$, one can analyze the network performance in the {\em ultrareliable} regime, where the target outage probability is close to $0$.\footnote{This regime is of high interest in wireless networks. The ultrareliable communication is a key requirement in modern wireless networks. For example, in vehicular networks, an extremely high reliability is mandatory.}

\subsubsection{Local Delay}
\label{sec:local_del_def}
The local delay is defined as the mean time (in terms of the number of time slots) until a packet is successfully delivered. Conditioned on the point process $\Phi$, the transmissions in different time slots are independent and succeed with probability $P_{\rm s}$. Hence the time until a successful packet delivery is a random variable with the geometric distribution and conditional mean $1/P_{\rm s}$ (after averaging over the fading and the channel access scheme). Consequently, the local delay $D$ is given by
\begin{align*}
D \triangleq \mathbb{E}^{!}_{o}\left(1/P_{\rm s}\right),
\end{align*}
where the expectation $\mathbb{E}^{!}_{o}(\cdot)$ corresponds to the probability $\mathbb{P}^{!}_{o}(\cdot)$ and is taken over the point process. Notice that the local delay is the $-1$st moment of the SIR MD. 

In summary, the density of reliable transmissions corresponds to reliable communication and the local delay corresponds to the latency as it yields the transmission delay.  
 
\subsection{Bandwidth Partitioning}
The total bandwidth $W$ is partitioned into $N$ orthogonal sub-bands of equal bandwidth. Thus the bandwidth of each sub-band is $W/N$. Each transmitter randomly selects a sub-band for transmission. Let $s_k(x) \in \lbrace 1, 2, \dotsc, N \rbrace$ denote the index of the sub-band that the transmitter at the location $x \in \Phi$ selects in $k$th time slot. Then the SIR at the typical receiver in $k$th time slot is given as
\begin{align*}
\mathsf{SIR}_k = \frac{h_{k,x_0}}{\sum_{x \in \Phi\setminus \lbrace x_0\rbrace}h_{k,x}\|x\|^{-\alpha}\boldsymbol{1}(s_k(x_0) = s_k(x))},
\end{align*} 
where $\boldsymbol{1}(\cdot)$ is the indicator function and $\|x\|$ is the distance of the transmitter $x$ from the origin.

 The data rate $R_N$ depends on the allocated bandwidth as $R_N = \frac{W}{N}\log_2(1 +\theta)$.
With BWP, there are two ways to send a packet of fixed size depending on whether one wishes to keep the data rate the same or not:
\begin{itemize}
\item[1)] {\em Adapt the SIR threshold}: In this approach, the data rate is maintained despite the reduction in the allocated bandwidth due to BWP, {\em i.e.}, we wish to have $R_N = R$.\footnote{Note that we have denoted $R$ to be the data rate when there is no BWP, {\em i.e.}, $R  =W\log_2(1+\theta)$.} To maintain the data rate, we have to adapt the SIR threshold accordingly. The required SIR threshold $\theta(N)$ can be obtained by inverting the rate expression as
\begin{align*}
\!R_N = R = \frac{W}{N}\log_2(1 + \theta(N)) \!\!\quad\!\! \Rightarrow \!\!\quad \theta(N)  = 2^{\frac{NR}{W}} - 1.
\end{align*}
A packet transmission still requires exactly one time slot. 
\item[2)] {\em Adapt the packet transmission time}: In this approach, the SIR threshold $\theta$ is kept unchanged irrespective of the value of $N$. Hence the data rate $R_N = \frac{W}{N}\log_2(1 + \theta)$ is reduced by $N$ compared to the case with no BWP, {\em i.e.},
\begin{align*}
R_N = R/N \quad \Rightarrow \quad \theta = 2^{R/W} - 1.
\end{align*}
This increases the time required to transmit a packet. Since a packet transmission requires one time slot when the allocated bandwidth is $W$, it needs $N$ time slots when the bandwidth $W$ is split into $N$ sub-bands.
\end{itemize}


\section{Density of Reliable Transmissions}
\subsection{Adaptive SIR Threshold Approach}
\label{sec:ASTA_dens}
As defined in~\eqref{eq:dens_suc}, the SIR MD framework can directly be used to calculate the density of reliable transmissions. But the direct calculation of the SIR MD is infeasible. Hence we take a detour where we first calculate the $b$th moment of the conditional success probability $P_{\rm s}$. 
\begin{theorem}[$b$th moment of $P_{\rm s}$]
\label{thm:Mb}
For the adaptive SIR threshold approach with $N$ sub-bands, the $b$-th moment $M_b$ of $P_{\rm s}$ is
\begin{align}
\!\!M_b(N) = \exp\!\left(\!\!-b \lambda  C\:_2F_1(1-b, 1-\delta; 2;1/N)\frac{\theta(N)^{\delta}}{N}\!\right)\!,
\label{eq:Mb_MD}
\end{align}
where $\delta \triangleq 2/\alpha$, $C = \pi \Gamma(1-\delta)\Gamma(1+\delta)$, $_2F_1(\cdot, \cdot; \cdot; \cdot)$ is the Gauss hypergeometric function, and $\theta(N) = 2^{\frac{NR}{W}} - 1$.
\end{theorem}
\begin{IEEEproof}
See Appendix~\ref{app:Mb_proof}.
\end{IEEEproof}
The exact SIR MD $\bar{F}_{P_{\rm s}}(x) = \mathbb{P}(P_{\rm s}(\theta) > x)$ can be calculated using the Gil-Pelaez theorem~\cite{gp_theorem} as 
\begin{align}
\bar{F}_{P_{\rm s}}(x)  =  \frac{1}{2}+\frac{1}{\pi} \int\limits_0^\infty \frac{\Im(e^{-jt \log x}M_{jt}(N))}{t}\mathrm{d}t ,
\label{eq:exact_meta}
\end{align}
where $\Im(\nu)$ is the imaginary part of $\nu \in \mathbb{C}$. Although the expression in \eqref{eq:exact_meta} calculates the SIR MD exactly, no useful insights can be obtained due to its complexity. Surprisingly, it is possible to get a very simple closed-form expression of the SIR MD in a practically important regime, which is the ultrareliable regime where the target reliability is close to $1$. In our notation, the ultrareliable regime corresponds to the one where $\varepsilon \to 0$, {\em i.e.}, the target outage probability goes to $0$. To obtain the SIR MD in a closed-form in the ultrareliable regime, we use the following two lemmas. 

The first lemma identifies an asymptotic property of the Gauss hypergeometric function.
\begin{lemma}
For $b \in \mathbb{R}$,
\begin{align*}
\:_2F_1(1-b, 1-\delta; 2;1/N) \sim \frac{\left(\frac{b}{N}\right)^{\delta-1}}{\Gamma(1+\delta)}, \quad b \to \infty.
\end{align*}
\label{lem:hyper_geom}
\end{lemma}\vspace*{-3mm}
The second lemma is borrowed from \cite[Theorem~1]{voss_2009}, which basically states the de Bruijn's Tauberian theorem.
\begin{lemma}
For a non-negative random variable $X$, the Laplace transform $\mathbb{E}[\exp(-tX)] ~ \sim \exp(p t^{u})$ for $t \to \infty$ is equivalent to the cumulative distribution function $\mathbb{P}(X \leq \varepsilon) \sim \exp(q/ \varepsilon^{v})$ for $\varepsilon \to 0$, when $1/u = 1/v + 1$ (for $u \in (0,1)$ and $v > 0$), and the constants $p$ and $q$ are related as $\lvert up\rvert^{1/u} = \lvert vq \rvert^{1/v}$.
\label{lem:bruijn}
\end{lemma}
The following theorem obtains the SIR MD in a closed-form as $\varepsilon \to 0$.
\begin{theorem}[]
\label{th:asym}
As $\varepsilon \to 0$, the SIR MD is
\begin{equation}
\bar{F}(\theta(N), \varepsilon) \sim \exp\left(-C_\delta \left(\frac{\theta(N)}{N\varepsilon}\right)^{\frac{\delta}{1-\delta}}\lambda^{\frac{1}{1-\delta}}\right), 
\label{eq:MD_asym}
\end{equation}
where $C_\delta \triangleq \frac{\left( \pi \delta \Gamma(1-\delta)   \right)^{1/(1-\delta)}}{\delta/(1-\delta)}$.
\label{thm:ultra_rel}
\end{theorem}
\begin{IEEEproof}
See Appendix~\ref{app:ultra_rel}.

\end{IEEEproof}

Consequently, in the ultrareliable regime, the density of reliable transmissions $\lambda_{\rm s}$ can be calculated as
\begin{align*}
\lambda_{\rm s} \sim \lambda \exp\left(-C_\delta \left(\frac{\theta(N)}{N\varepsilon}\right)^{\frac{\delta}{1-\delta}}\lambda^{\frac{1}{1-\delta}}\right), \quad \varepsilon \to 0,
\end{align*}
which reveals that the intensity $\lambda$ has two competing effects on $\lambda_{\rm s}$. To maximize $\lambda_{\rm s}$, we need to obtain the optimal pair $(\lambda, N)$. The following theorem gives the maximum density of reliable transmissions in the ultrareliable regime.
\begin{theorem}
\label{thm:max_dens}
As $\varepsilon \to 0$, the maximum density of reliable transmissions $\mathsf{S}$ is 
\begin{align}
\mathsf{S} \triangleq \sup_{\lambda, N} \lambda_{\rm s} \sim \left(\frac{\varepsilon}{\theta_1}\right)^{\delta}\frac{1}{\pi e^{1-\delta}\delta^\delta\Gamma(1-\delta)}.
\label{eq:dens_max}
\end{align}
It is achieved at $\lambda = \frac{\varepsilon^{\delta}}{\pi \delta^\delta\Gamma(1-\delta)\theta_1^{\delta}}$ and $N = 1$, where $\theta_1 = 2^{R/W}-1$.
\end{theorem}
\begin{IEEEproof}
We relax $N >0$ to be continuous and take the nearest integer as the optimal value. The objective at hand is:
\begin{align*}
f(\lambda, N) = \lambda \exp\left(-B \left(\frac{\theta(N)}{N}\right)^{\frac{\delta}{1-\delta}}\lambda^{\frac{1}{1-\delta}}\right),
\end{align*}
where $B = \frac{C_{\delta}}{\varepsilon^{\delta/(1-\delta)}}$ and $\theta(N) = 2^{NR/W}-1$. We obtain the critical point $\lambda_0(N)$ by letting $\frac{\partial f(\lambda, N)}{\partial \lambda} = 0$, {\em i.e.},
\begin{align*}
\underbrace{\exp\!\!\left(\!\!-B \!\left(\!\frac{\theta(N)}{N}\!\right)^{\!\frac{\delta}{1-\delta}}\!\lambda^{\frac{1}{1-\delta}}\!\!\right)}_{> 0}\!\!\!\left[\!1 - \frac{B}{1-\delta}\!\!\left(\frac{\theta(N)}{N}\right)^{\frac{\delta}{1-\delta}}\!\!\lambda^{\frac{1}{1-\delta}}\!\right] = 0.
\end{align*}
It follows that
\begin{align*}
\lambda_0(N) = \left(\frac{\theta(N)}{N}\right)^{-\delta} \left(\frac{1-\delta}{B}\right)^{1-\delta}.
\end{align*}
Since the objective function is quasiconcave for any given $N > 0$, we have
\begin{align*}
\mathsf{S} &\sim \sup_{N} f(\lambda_0(N), N), \quad \varepsilon \to 0 \nonumber \\
& = \left(\frac{1-\delta}{eB}\right)^{1-\delta}~\sup_{N}~ \left(\frac{\theta(N)}{N}\right)^{-\delta}.
\end{align*}
From the Taylor series of $2^{NR/W}$, it can be shown that $\theta(N)/N = (2^{NR/W}-1)/N$ increases strictly  monotonically with $N$. Hence $N = 1$ maximizes the density of reliable transmissions. Consequently, we have $\mathsf{S}$ as in~\eqref{eq:dens_max}, which is achieved at $\lambda = \frac{\varepsilon^{\delta}}{\pi \delta^\delta\Gamma(1-\delta)\theta_1^{\delta}}$.
\end{IEEEproof}

\begin{figure}
\centering
\includegraphics[width = 8.1cm, height = 6cm]{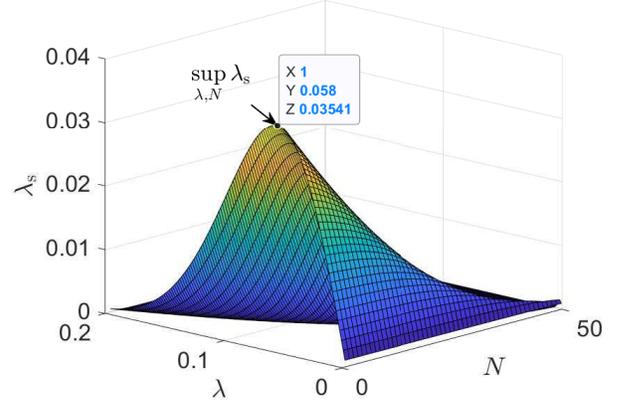}\vspace*{-1mm}
\caption{The $3$D plot of the density of reliable transmissions against $\lambda$ and $N$ for the adaptive SIR threshold approach. The maximum density of reliable transmissions is achieved at $\lambda = 0.0584$ and $N = 1$ and is equal to $0.0354$. 
$R = 0.1$, $W = 1$, $\alpha = 4$, and $\varepsilon = 0.01$.}
\label{fig:max_dens_suc}\vspace*{-2mm}
\end{figure}

\begin{remark}
For the adaptive SIR threshold approach, the density of reliable transmissions is maximized when there is no bandwidth partitioning. This is because the exponential rise in the SIR threshold with $N$ to maintain the data rate (which reduces the reliability) dominates the advantage of smaller interference due to bandwidth partitioning.
\end{remark}
Fig.~\ref{fig:max_dens_suc} shows the behavior of the density of reliable transmissions $\lambda_{\rm s}$ against $\lambda$ and $N$.

\subsection{Adaptive Transmission Time Approach}

In this approach, the SIR threshold does not depend on the number of sub-bands $N$. Hence the $b$th moment of $P_{\rm s}$ and the SIR MD in the ultrareliable regime can be calculated using \eqref{eq:Mb_MD} and \eqref{eq:MD_asym}, respectively, by simply replacing $\theta(N) = 2^{NR/W} - 1$  with $\theta = 2^{R/W} - 1$.  

Following the steps discussed for the adaptive SIR threshold approach (Section~\ref{sec:ASTA_dens}), in the ultrareliable regime, the maximum density of reliable transmissions behaves as
\begin{align*}
\mathsf{S} \to \infty, \quad \varepsilon \to 0,
\end{align*}
which is achieved as $N \to \infty$ and $\lambda \to \infty$. We observe this behavior because the fraction of reliable links, {\em i.e.}, the SIR MD, is maximized as $N \to \infty$ since the SIR threshold does not depend on $N$ unlike the adaptive SIR threshold approach and the advantage of smaller interference due to BWP helps increase the reliability. This behavior is in contrast with that in the adaptive SIR threshold approach where $N = 1$ (the other extreme end of BWP) maximizes $\lambda_{\rm s}$.

\section{The Local Delay}
\subsection{Adaptive SIR Threshold Approach}
As discussed in~Section~\ref{sec:local_del_def}, the local delay is the $-1$st moment of the SIR MD. Hence the local delay $D$ can be obtained by directly substituting $b = -1$ in \eqref{eq:Mb_MD} as
\begin{align}
D(N) = M_{-1}(N) =  \exp\left(\lambda C\left(\frac{\theta(N)}{N}\right)^{\delta}(N-1)^{-(1-\delta)}\right),
\label{eq:loc_del}
\end{align}
where $C =  \pi \Gamma(1-\delta)\Gamma(1+\delta)$. 

\begin{remark}
For $N = 1$ which maximizes the density of reliable transmissions, the local delay is infinite, i.e., a successful packet transmission requires infinite time slots. This is because all transmitters share the same frequency band and there are interferers close enough to the typical receiver to result in unsuccessful packet transmissions. Hence, bandwidth partitioning helps achieve a finite local delay.
\end{remark}
From \eqref{eq:loc_del}, we also observe that the local delay is infinite as $N \to \infty$. Hence there exists a finite optimum $N$ that minimizes the local delay, which we obtain in the following theorem.
\begin{theorem}
\label{thm:min_local}
Let $N_0 > 1$ be the unique solution of
\begin{align}
(a\delta N(N-1)\log(2) - N +\delta)2^{aN} + N -\delta = 0,
\label{eq:loc_fixed_eq}
\end{align}
where $a = R/W$. Also, let $\mathsf{Round}(x)$ denote the nearest integer if $2 < x < \infty$ and equals $2$ when $1 < x \leq 2$.
Then $N^* = \mathsf{Round}(N_0)$ is the optimum number of sub-bands that minimizes the local delay.
\end{theorem}
\begin{IEEEproof}
To obtain the optimum number of sub-bands $N^*$, we relax $N$ to be continuous. Then $N^*$ is its nearest integer. 

From~\eqref{eq:loc_del}, we can see that $N = 1$ leads to the infinite local delay. Hence $N^*$ is at least $2$. For $N > 1$, we take the derivative of $D(N)$ given in~\eqref{eq:loc_del} w.r.t. $N$, which is
\begin{align*}
D'(N) = g(N) \left(\frac{2^{aN}-1}{N}\right)^{\delta}(N-1)^{\delta-1},
\end{align*}
where $g(N) = \frac{(a\delta N(N-1)\log(2) - N +\delta)2^{aN} + N -\delta}{N(N-1)(2^{aN}-1)}$. It can be observed that $g(N)$ strictly monotonically increases with $N$, which implies that there exists only one optimal value $N^*$ of $N$ that satisfies $D'(N^*) = 0$. Consequently, $N^*$ is obtained from solving $g(N) = 0$, {\em i.e.}, $(a\delta N(N-1)\log(2) - N +\delta)2^{aN} + N -\delta = 0$. Since $N$ is a positive integer, we take the nearest integer greater than $1$ as the optimum $N$.
\end{IEEEproof}
Although it is not possible to obtain a closed-form solution to \eqref{eq:loc_fixed_eq}, it can easily be obtained numerically. 
\subsection{Adaptive Transmission Time Approach}
For the local delay, this approach has been studied in~\cite{Zhong2014}. We briefly discuss it here for the sake of completeness. In this approach, the expression of the local delay is obtained by replacing $\theta(N)$ in \eqref{eq:loc_del} by $\theta = 2^{R/W} - 1$ and multiplying the complete expression by $N$ since a packet transmission (irrespective of whether it is successful or not) requires $N$ time slots. Thus the local delay can be expressed as
\begin{align*}
D(N) = N \exp\left(\lambda C\left(\theta/N\right)^{\delta}(N-1)^{-(1-\delta)}\right),
\end{align*}
which is the same as (5) of \cite{Zhong2014}. 
\begin{remark}
For this approach as well, $N = 1$ and $N \to \infty$ make the local delay infinite. Hence there exists a finite $N$ that minimizes the local delay.
\end{remark} 
As shown in~\cite[Theorem 5]{Zhong2014}, the optimum number of sub-bands $N^*$ is the unique solution of the following fixed point equation
\begin{align}
\frac{1}{N}\left(C \theta^\delta\left(\frac{N}{N-1}\right)^{1-\delta}\frac{N-\delta}{N-1}\right) = 1.
\label{eq:loc_fixed_eq1}
\end{align}
It has also been shown in~\cite{Zhong2014} that $N^*$ is tightly bounded as
\begin{align*}
N^*\in [\floor{\lambda C\theta^{\delta}}, \ceil{\lambda C\theta^{\delta}} + 2].
\end{align*}

\begin{figure}
\centering
\includegraphics[width = 8.1cm, height = 5.6cm]{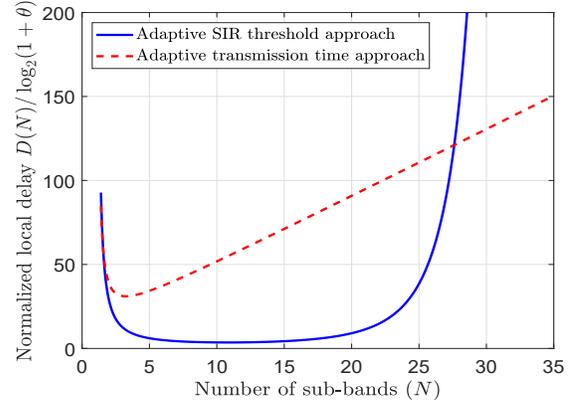}
\caption{The normalized local delay $\frac{D(N)}{\log_2(1+\theta)}$ against the number of sub-bands ($N$). For the adaptive SIR threshold approach, $\theta = 2^{NR/W}-1$. For the adaptive transmission time approach, $\theta = 2^{R/W}-1$. $R = 0.25$, $W = 1$, $\alpha = 3$, and $\lambda = 1$.}
\label{fig:comp_local_delay}\vspace*{-4mm}
\end{figure}

Fig.~\ref{fig:comp_local_delay} compares the delay performance for the adaptive SIR threshold and the adaptive time approaches. Specifically, Fig.~\ref{fig:comp_local_delay} plots the normalized local delay given by $\frac{D(N)}{\log_2(1+\theta)}$ against the number of sub-bands $N$.\footnote{Note that the local delay is measured in number of time slots, and the time-slot duration is proportional to $\frac{1}{\log_2(1+\theta)}$ since the packet size is fixed and the spectral efficiency is proportional to $\log_2(1+\theta)$. Hence, to compare actual delays for different SIR thresholds $\theta$, we need to normalize the local delay $D(N)$ by $\log_2(1+\theta)$.} It shows that there exists a finite $N$ for which the delay is minimized. Hence, BWP helps reduce the local delay. Also, we observe that when $N$ is small to moderate, the adaptive SIR threshold approach experiences a smaller delay. But as $N$ increases, the adaptive transmission time approach has a better delay performance since the exponential rise in the SIR requirement for the adaptive SIR threshold approach reduces the success probability (reliability), which causes frequent failed transmissions and thus requires more number of time slots to successfully transmit a packet. This negative effect in the adaptive SIR threshold approach dominates the negative aspect of the adaptive transmission time approach that each packet transmission (successful or not) requires $N$ time slots. 

Table~\ref{tab:1} summarizes the optimum number of sub-bands $N^*$ corresponding to the maximum density of reliable transmissions in the ultrareliable regime and the minimum local delay.

\section{Conclusions}
The answer to the question ``Does BWP help improve the reliability and delay performance in wireless networks?" is both \textit{yes} and \textit{no}. For the adaptive SIR threshold approach, BWP is not helpful if one wishes to maximize the density of reliable transmissions in the ultrareliable regime. On the other hand, ``extreme" BWP, {\em i.e.}, $N \to \infty$, maximizes the density of reliable transmissions for the adaptive transmission time approach. In fact, both cases of $N = 1$ and $N \to \infty$ make the local delay infinite for both the  approaches. The optimum $N$ that minimizes the local delay takes none of the extreme values of $N$ ({\em i.e.}, $N = 1$ or $N \to \infty$) and lies somewhere in between depending on the system parameters such as the path-loss exponent and the intensity of the underlying point process. Hence, one needs to select $N$ appropriately to maximize the density of reliable transmissions in the ultrareliable regime while keeping the local delay below a threshold. 
From a broader perspective, since the reliability and the delay are the components of the ultrareliable low-latency communication (URLLC), the results in this paper may help understand the effect of BWP on the URLLC dynamics.
\begin{table}
\caption{Optimal number of sub-bands $N^*$}
\begin{center}
\begin{tabular}{|l|c|c|}
\hline
\diagbox{Approach}{Metric} & \!\!\!$\displaystyle\sup_{\lambda, N} \lambda_{\rm s}$, $\varepsilon \to 0$\!\!\! & $D(N)$ \\
\hline
Adaptive SIR &  &Solution to \eqref{eq:loc_fixed_eq},\\ threshold &  $1$  & $N^* \geq 2$\\ \hline
Adaptive transmission &&Solution to \eqref{eq:loc_fixed_eq1},\\ time &  $ \to \infty$ & \!\!\!$N^* \in [\floor{\lambda C\theta^{\delta}}, \ceil{\lambda C\theta^{\delta}} + 2]$\!\!\! \\
\hline
\end{tabular}
\label{tab:1}
\end{center}\vspace*{-4mm}
\end{table}

\section*{Acknowledgment}
The author would like to thank Martin Haenggi and Fran\c cois Baccelli for the initial discussion on this problem. 

This work has received funding from the European Research Council (ERC) under the European Union's Horizon 2020 research and innovation programme grant agreement number 788851.
\appendices
\section{Proof of Thm.~\ref{thm:Mb}}
\label{app:Mb_proof}
In the $k$th time slot, the success probability $P_{\rm s}$ conditioned on $\Phi$ is 
\begin{align*}
P_{\rm s} &= \mathbb{P}(\mathsf{SIR}_k > \theta(N) \mid \Phi) = \mathbb{P}\left(\frac{h_{k,x_0}}{I_k} > \theta(N) \mid \Phi\right),
\end{align*}
where $I_k = \sum_{x \in \Phi\setminus \lbrace x_0\rbrace}h_{k,x}\|x\|^{-\alpha}\boldsymbol{1}(s_k(x_0) = s_k(x))$. By averaging over the fading $h_{k,x_0}$ on the desired link, we have
\begin{align*}
P_{\rm s} &= \mathbb{E}(e^{-\theta(N) I_k} \mid \Phi) \\
 &\!\!\!\!= \hspace*{-2mm}\!\prod_{x \in \Phi\setminus \lbrace x_0\rbrace}\hspace*{-4mm}\mathbb{E}\!\left(\exp\left(-\theta(N) h_{k,x}\|x\|^{-\alpha}\boldsymbol{1}(s_k(x_0) = s_k(x))\right) \mid \Phi\right)\!.
\end{align*}
Averaging over the random sub-band selection and the fading on interfering links, it follows that
\begin{align}
P_{\rm s} = \prod_{x \in \Phi\setminus \lbrace x_0\rbrace} \left(\frac{1}{N}\frac{1}{1 + \theta(N)  \|x\|^{-\alpha}} + \frac{N-1}{N}\right).
\end{align}
The $b$th moment $M_b = \mathbb{E}(P_{\rm s}^{b})$ of $P_{\rm s}$ is given by
\begin{align}
\!M_b 
& \!\overset{(\mathrm{a})}{=} \!\exp\!\!\left(\!\!-2 \pi \lambda\!\! \int_{0}^{\infty}\!\!\left[\!1 - \!\left(\!\frac{1}{N(1 + \theta(N)  r^{-\alpha})} + \frac{N-1}{N}\!\!\right)^b\!\right]\!\!r\mathrm{d}r\!\!\right) \nonumber \\
& = \exp\left(-b \lambda C \:_2F_1(1-b, 1-\delta; 2;1/N)\frac{\theta(N)^{\delta}}{N}\right), \nonumber
\end{align}
where $(\mathrm{a})$ follows from the probability generating functional (PGFL) of the PPP~\cite[Chapter 4]{martin_book}.

\section{Proof. of Thm.~\ref{thm:ultra_rel}}
\label{app:ultra_rel}

The Laplace transform of $X = -\log(P_{\rm s})$ is $\mathbb{E}(\exp(-tX))$ = $\mathbb{E}(P_{\rm s}^t)$ = $M_{t}$, which is the $t$th moment of $P_{\rm s}$. From \eqref{eq:Mb_MD}, Lemma~\ref{lem:hyper_geom}, and Lemma~\ref{lem:bruijn}, it follows that
\begin{equation}
M_t \sim \exp\left(- \frac{\lambda C (\theta(N)/N)^\delta t^\delta}{ \Gamma(1+\delta)}\right), \quad |t| \to \infty.
\label{eq:MD_asym_int}
\end{equation}
Also,
\begin{align*}
\mathbb{P}(X \leq \varepsilon) &= \mathbb{P}(P_{\rm{s}} \geq \exp(-\varepsilon)) \nonumber \\
&\overset{(\mathrm{a})}{\sim} \mathbb{P}(P_{\rm{s}} \geq 1-\epsilon), \quad \varepsilon \to 0 \nonumber \\
 &\overset{(\mathrm{b})}{=} \exp\left(-C_\delta \left(\frac{\theta(N)}{N\varepsilon}\right)^{\frac{\delta}{1-\delta}}\lambda^{\frac{1}{1-\delta}}\right),
\end{align*}
where $(\mathrm{a})$ is due to $\exp(-\varepsilon) \sim 1-\varepsilon$ as $\varepsilon \to 0$ and $(\mathrm{b})$ is due to Lemma~\ref{lem:bruijn}.

\bibliographystyle{ieeetr}
\bibliography{paper}

\end{document}